\documentclass[showpacs,twocolumn,floats,superscriptaddress,prl,eqsecnum]{revtex4}
\usepackage{graphicx}

\begin{document}
\title{Low field phase diagram of spin-Hall effect in the mesoscopic regime}
\author{Zhenhua Qiao}
\affiliation{Department of Physics, The University of Hong Kong,
Hong Kong, China}
\author{Wei Ren$^*$}
\affiliation{Department of Physics, The University of Hong Kong,
Hong Kong, China}
\author{Jian Wang$^\dag$}
\affiliation{Department of Physics and the center of theoretical and
computational physics, The University of Hong Kong, Hong Kong,
China}
\author{Hong Guo}
\affiliation{Center for the Physics of Materials \& Department of
Physics, McGill University, Montreal, PQ, Canada}

\begin{abstract}

When a mesoscopic two dimensional four-terminal Hall cross-bar with
Rashba and/or Dresselhaus spin-orbit interaction (SOI) is subjected
to a perpendicular uniform magnetic field $B$, both integer quantum
Hall effect (IQHE) and mesoscopic spin-Hall effect (MSHE) may exist
when disorder strength $W$ in the sample is weak. We have calculated
the low field ``phase diagram" of MSHE in the $(B,W)$ plane for
disordered samples in the IQHE regime. For weak disorder, MSHE
conductance $G_{sH}$ and its fluctuations $rms(G_{SH})$ vanish
identically on even numbered IQHE plateaus, they have finite values
on those odd numbered plateaus induced by SOI, and they have values
$G_{SH}=1/2$ and $rms(G_{SH})=0$ on those odd numbered plateaus
induced by Zeeman energy. For moderate disorder, the system crosses
over into a regime where both $G_{sH}$ and $rms(G_{SH})$ are finite.
A larger disorder drives the system into a chaotic regime where
$G_{sH}=0$ while $rms(G_{SH})$ is finite. Finally at large disorder
both $G_{sH}$ and $rms(G_{SH})$ vanish. We present the physics
behind this ``phase diagram".
\end{abstract}
\pacs{
71.70.Ej,  
72.15.Rn,  
72.25.-b   
}
\maketitle

Many recent papers have been devoted to the physics of spin-Hall
effect\cite{hirsch} and a particular focus is the {\it intrinsic}
spin-Hall generated in non-magnetic samples by spin-orbital
interaction (SOI)\cite{murakami,sinova}. So far, several
experimental papers have reported observations of spin-Hall effect
in compound semiconductors and other systems\cite{exp1}.
Theoretically, it has been shown that for two dimensional (2D)
samples in the clean limit, the Rashba SOI generates a spin-Hall
conductivity having a universal value of $e/8\pi$\cite{sinova}. The
presence of weak disorder destroys spin-Hall effect in large
samples\cite{inoue,mishchenko}. In particular, a consensus appears
to have been reached in the literature that spin-Hall effect in
disordered samples generated by linear Rashba SOI vanishes at the
thermodynamical limit\cite{mishchenko,nagaosa1,nomura}.

For \emph{mesoscopic} samples, numerical studies have provided
evidence that the \emph{mesoscopic spin-Hall effect} (MSHE) can
survive weak disorder\cite{hank,sheng1,nikolic,weng1}. For a
four-probe disordered sample, MSHE conductance $G_{SH}$ and its
fluctuations $rms(G_{SH})$ have been calculated for both linear
Rashba and Dresselhaus SO interactions\cite{sheng1,ren1}. It was
found\cite{ren1} that when the system is in the diffusive regime,
the fluctuations $rms(G_{SH})$ take a universal value with the same
order of magnitude as the average $G_{SH}$ itself, and is
independent of the system size $L$, the disorder strength $W$, the
electron Fermi energy and the SO interaction strength.

The situation becomes very interesting and more complicated when a
perpendicular uniform external magnetic field $B$ is applied to the
2D sample\cite{shen1}. In this case, $G_{SH}$ and $rms(G_{SH})$
become functions of $B$. Most importantly, a magnetic field $B$ can
produce edge-states which are responsible for the integer quantum
Hall effect (IQHE). Similar to the well known studies of the global
phase diagram of quantum Hall effect\cite{kivelson}, it will be very
useful to map out the low field ``phase diagram" of MSHE in terms of
the field strength $B$ and the disorder strength $W$. Such a diagram
allows one to clearly understand the role played by the edge-states
and disorder. It is the purpose of this work to present this MSHE
``phase diagram" for four-probe 2D disordered mesoscopic samples
with linear Rashba and/or Dresselhaus SO interactions.

\begin{figure}
\includegraphics[width=6.3cm,angle=270]{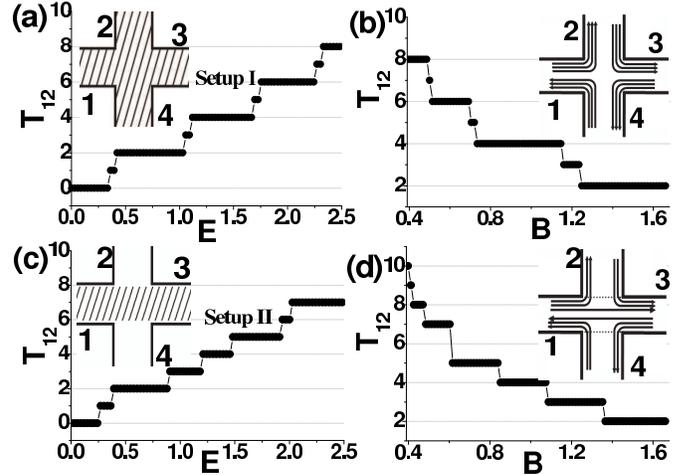}
\caption{Transmission coefficient $T_{12}$ versus $E$ or $B(T)$. For
setup-I: (a) and (b). For setup-II: (c) and (d). Inset of (a):
schematic plot of the setup-I; inset of (b): the corresponding flow
of edge states. Inset of (c): schematic plot of the setup-II; inset
of (d): the corresponding flow of edge states. } \label{fig1}
\end{figure}

Here we put ``phase diagram" in quotes because the physics we study
is mesoscopic, namely for samples in the coherent diffusive regime
characterized by the relation between relevant length scales,
$l<L<\xi$. Here $L$ is the linear sample size, $l$ the elastic mean
free path and $\xi$ the phase coherence length. As such, the
``phases" in the ``phase diagram" are states with zero or finite
values of $G_{SH}$ and $rms(G_{SH})$, and no phase transitions are
implied between these states. In particular, we found that with low
disorder when IQHE is well established, both $G_{SH}$ and
$rms(G_{SH})$ are zero identically on the \emph{even} numbered IQHE
plateaus, while they take finite values on the SOI dominant
\emph{odd} numbered IQHE plateaus. For Zeeman dominant \emph{odd}
numbered IQHE plateaus, $G_{sH}=1/2$ and its fluctuation vanishes.
As the disorder is increased, both $G_{SH}$ and $rms(G_{SH})$ become
nonzero when any edge-state is destroyed by the disorder in any IQHE
plateau. Further increase of disorder brings the system to a
``chaotic" regime where $G_{SH}=0$ while $rms(G_{SH})\neq 0$,
finally at even larger disorder both $G_{SH}$ and $rms(G_{SH})$
vanish. These behaviors are organized in the low field phase diagram
which we determine in the rest of the paper.

We consider a 2D four-probe device schematically shown in the inset
of Fig.\ref{fig1}c (call it setup-II). A MSHE conductance $G_{SH}$
is measured\cite{sheng1} across probes labeled $2,4$ when a small
voltage bias is applied across probes $1$ to $3$ so that a current
flows between them. $G_{SH}$ can be measured the same way when there
is a uniform external magnetic field $B$ which exists everywhere
including inside the leads. $G_{SH}$ is theoretically calculated
from spin-current defined as $I_s\equiv
\hbar/2(I_{\uparrow}-I_{\downarrow})$ where
$I_{\uparrow,\downarrow}$ are contributions from the two spin
channels. Note that the definition of $I_s$ is, in fact, in debate
for regions where SO interaction exists\cite{niu1,nagaosa1}. To
avoid this ambiguity we assume that in our device the SO interaction
only exists in the shaded region (setup-II in Fig.\ref{fig1}c),
namely in leads $1,3$ and in the central scattering region, but does
not exist in leads $2,4$ where we measure spin-current. This way,
$I_s$ is well defined as above. For discussion purposes, we have
also considered a device (setup-I, inset of Fig.\ref{fig1}a) where
SO interaction is present everywhere including inside leads $2,4$.

In the presence of linear Rashba interaction $\alpha_{so} {\bf z}\cdot
({\bf \sigma} \times {\bar {\bf k}})$ with ${\bar {\bf k}}={\bf k}+(e/\hbar c)
{\bf A}$, the Hamiltonian of the four-probe device is:
\begin{eqnarray}
H&=& \sum_{nm\sigma}\epsilon_{nm} c^\dagger_{nm\sigma} c_{nm\sigma}
+ g_s\sum_{nm\sigma \sigma'} c^\dagger_{nm\sigma} (\sigma \cdot
{\bf B})_{\sigma \sigma'} c_{nm\sigma'}
\nonumber \\
&-&t\sum_{nm\sigma}[c^\dagger_{n+1,m\sigma} c_{nm\sigma} e^{-im\eta}
+c^\dagger_{n,m-1\sigma} c_{nm\sigma} +h.c.] \nonumber \\
&-&t_{so}\sum_{nm\sigma\sigma'} [c^\dagger_{n,m+1\sigma}
(i\sigma_x)_{\sigma \sigma'} c_{nm\sigma'} \nonumber \\
&-&c^\dagger_{n+1,m\sigma} (i\sigma_y)_{\sigma \sigma'} c_{nm\sigma'}
e^{-im\eta} +h.c.]
\label{eq1}
\end{eqnarray}
where $c^\dag_{nm\sigma}$ is the creation operator for an electron with spin
$\sigma$ on site $(n,m)$, $\epsilon_{nm\sigma}=4t$ is the on-site energy,
$t=\hbar^2/2\mu a^2$ is the hopping energy and $t_{so} = \alpha_{so}/2a$ is the
effective Rashba spin-orbit coupling, $g_s=(1/2)g\mu_B$ (with $g=4$) is
the Lande g factor. Here $\eta=\hbar \omega_c/2t$ and $\omega_c\equiv
eB/\mu c$ is the cyclotron frequency. Throughout this paper, we use $t$
as the unit of energy. For $L=40a=1\mu m$, $t=1.5\times 10^{-3} {\rm eV}$,
and $t_{so}=0.2t$ corresponds to $\alpha_{so}=9
\times 10^{-12} {\rm eV.m}$\cite{shen1}. We choose
${\bf A} = (-By,0,0)$ so that the system has translational symmetry along
x-direction (from lead 1 to lead 3). Static Anderson-type disorder is added
to $\epsilon_{i}$ with a uniform distribution in the interval $[-W/2,W/2]$
where $W$ characterizes the strength of the disorder. The spin Hall conductance
$G_{sH}$ is calculated from the Landauer-Buttiker formula\cite{hank}
\begin{equation}
G_{sH}=(e/8\pi)[(T_{2\uparrow,1}-T_{2\downarrow,1})-(T_{2\uparrow,3}
-T_{2\downarrow,3})]
\label{GsH}
\end{equation}
where transmission coefficient is given by $T_{2\sigma,1} ={\rm
Tr}(\Gamma_{2\sigma} G^r \Gamma_1 G^a)$. Here $G^{r,a}$ are the
retarded and advanced Green's functions of central disordered region
of the device which we evaluate numerically. The quantities
$\Gamma_{i\sigma}$ are the line width functions describing coupling
of the leads to the scattering region and are obtained by
calculating self-energies due to the semi-infinite leads using a
transfer matrices method\cite{lopez84}. The spin-Hall conductance
fluctuation is defined as $\text{rms}(G_{sH})\equiv
\sqrt{\left\langle G_{sH}^{2}\right\rangle -\left\langle
G_{sH}\right\rangle ^{2}}$, where $\left\langle
{\cdots}\right\rangle $ denotes averaging over an ensemble of
samples with different disorder configurations of the same strength
$W$. The devices in Fig.\ref{fig1} have $L\times L$ central square,
and without losing generality we fixed $L=40$ grid points in our
numerics.

Before presenting the numerically determined ``phase diagram" for
the physics of MSHE using setup-II, let's first discuss the general
physics of spin-Hall current. For this purpose we use setup-I where
the SOI is everywhere so that the discussion is simpler. We first
examine the spin-Hall ``phase diagram" in the absence of SOI. In a
magnetic field, edge-states are formed. Fig.\ref{fig1}a,b shows
transmission coefficient $T_{12}$ for setup-I, which measures the
number of edge-states, versus Fermi energy $E$ or magnetic field
$B$. We observe that $T_{12}$, or the number of edge-states,
increases as $E$ for a fixed $B$ and it decreases as $B$ is
increased for a fixed $E$. Notice that the number of edge-states $N$
can be either even or odd. The odd $N$ region in $E$ or $B$ is very
narrow and is due to the Zeeman splitting that breaks the spin
degeneracy. When $N$ is even, spin-Hall current vanishes because all
the edge-states are fully polarized with half of them pointing to
one direction (say spin-up) and the other half pointing to opposite
direction (spin-down). When $N$ is odd, the spin-Hall conductance is
$1/2$. At weak disorder when all the edge-states survive, we
therefore conclude that $G_{sH}=0$ when $N$ is even and $G_{sH}
=1/2$ when $N$ is odd. Furthermore, it is useful to examine
fluctuations of the spin-Hall conductance $rms(G_{sH})$ for these
edge-states: we expect no fluctuations for all edge-states. As
disorder strength $W$ is increased, we reach a point where at least
one of the edge-states is destroyed and the system is in
a spin-Hall liquid state characterized by $G_{sH} \neq 0$ and
$rms(G_{sH}) \neq 0$ for any $N$. Further increasing $W$, we expect
strong scattering to bring the system into a chaotic state of MSHE,
characterized by $G_{sH}=0$ and $rms(G_{sH}) \neq 0$. At even larger
$W$, the system enters a spin-Hall insulator state where
$G_{sH}=rms(G_{sH})=0$.

Next, we turn on the SOI and discuss its effect on the ``phase
diagram". Fig.\ref{fig1}c,d show transmission coefficient $T_{12}$
for setup-I versus $E$ or $B$ for a fixed Rashba SOI $t_{so}=0.2$.
We observe that the behavior of $T_{12}$ is similar to that of
Fig.\ref{fig1}a,b except that the region of odd $N$ is now much
larger. When $N$ is even, spin-Hall current vanishes as before. In
the region of $B$ when $N$ is odd, two cases occur due to the
competition between SOI which tends to randomize the spin
polarization and the Zeeman energy which favors spin polarization
along a fixed direction. If Zeeman energy is large enough, then
$G_{sH}=1/2$ as before with $rms(G_{sH})=0$ while if SOI dominates
then there is at least one edge-state that has both spin-up and down
components: our numerical results show that the composition depends
on systems parameters. As a result, there is a net spin-Hall current
when $N$ is odd. This discussion becomes clearer when we examine
setup-II where the spin direction can be defined. At weak disorder
when all the edge-states survive, we have the same conclusion as
before, \emph{i.e.} $G_{sH}=0$ when $N$ is even and $G_{sH} \neq 0$
when $N$ is odd. We expect no fluctuations for even $N$ and for
those odd $N$ edge-states with $G_{sH}=1/2$, but finite fluctuations
for the rest of odd $N$ edge-states. Hence, at weak disorder, we
have a ``phase" of edge-state induced spin-Hall insulator with even
$N$ characterized by $G_{sH}= rms(G_{sH})=0$; a ``phase" of
edge-state induced spin-Hall liquid (but fluctuationless and Zeeman
dominant) with odd $N$ characterized by $G_{sH} =1/2$ and
$rms(G_{sH}) =0$; and finally a ``phase" of edge-state induced
spin-Hall liquid (SOI dominant) with odd $N$ characterized by
$G_{sH} \neq 0$ and $rms(G_{sH}) \neq 0$. As we increase the
disorder strength, the ``phase diagram" evolves through three
regimes similar to the case when SOI is off: a spin-Hall liquid
regime, a chaotic regime, and a spin-Hall insulating regime.

The discussion in the last paragraph gives the entire expectation
for the low field MSHE ``phase diagram". The problem of this
discussion is that the spin-Hall current is not well defined in
regions where SO interaction exists\cite{niu1,nagaosa1} such as
setup-I of Fig.\ref{fig1}a. Therefore, in the rest of the work we
consider setup-II where SO interaction does not exist in leads $2,4$
so that spin-Hall current is well defined and measurable without
ambiguity. The extra complication of setup-II is that there is an
interface between spatial region with $t_{so}=0$ and that with
$t_{so} \neq 0$. This interface acts as a potential barrier causing
additional scattering of edge-states. In particular, at certain
energies one of the edge-states goes directly from lead 1 to lead 3
due to this interface scattering. Insets of Fig.1b and Fig.1d show
schematically the edge states for setup-I and II, respectively. In
the inset of Fig.1d, however, an edge-state is now transmitted
directly from lead 1 to lead 3 due to the interface scattering just
discussed. We have confirmed that this is a generic feature which
occurs at different Fermi energies. For a fixed Fermi energy, this
can also happen when $B$ is varied. In Fig.2b, we plot the $T_{12}$
for setup-I, and $T_{12}$, $T_{13}$ for setup-II, at $W=0$. We
observe that $N=$ odd edge-states are much easier to be scattered
while the $N=$ even edge-states are stable against interface
scattering. Therefore, the regions in the MSHE ``phase diagram"
where $N=$ even becomes larger for setup-II than for setup-I. For
instance, the magnetic field $B$ for the onset of $N=2$ edge-state
changes from 1.32T to 1.2T due to the interface scattering (for a
device with lead width $L=1\mu m$). We emphasis that except for this
extra complication of interface scattering in setup-II, the general
physics discussion of MSHE ``phase diagram" for setup-I in the last
paragraph, holds perfectly for setup-II.

\begin{figure}
\includegraphics[width=6.3cm,angle=270]{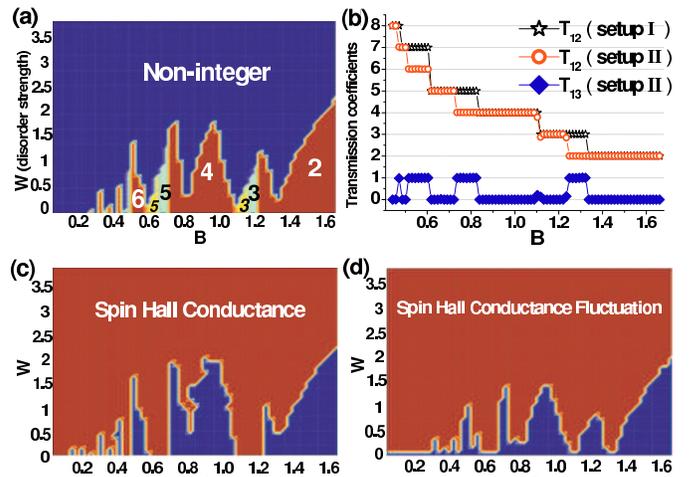}
\caption{(color online) (a). The edge state plateaus in $(B,W)$
plane. (b). The transmission coefficient $T_{12}$ for setup-I,
setup-II, as well as direct transmission coefficient $T_{13}$ as a
function of $B$ in the absence of disorder. (c). The spin-Hall
conductance in $(B,W)$ plane. (d). The spin-Hall conductance
fluctuation in $(B,W)$ plane. } \label{fig2}
\end{figure}

\begin{figure}
\includegraphics[width=6.5cm,angle=270]{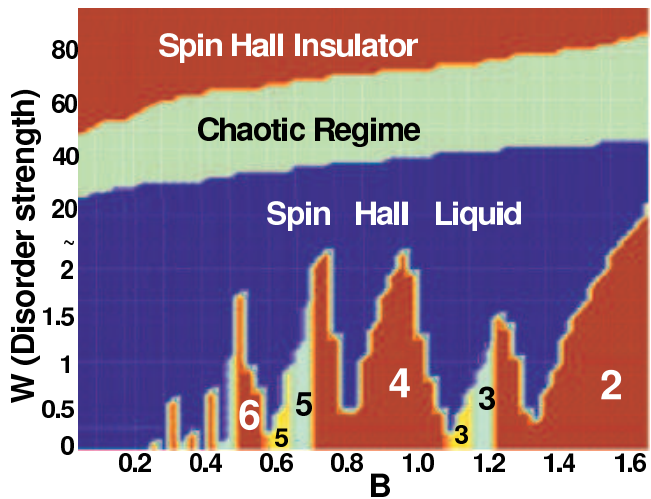}
\caption{(color online) The low field ``phase diagram" of mesoscopic
spin-Hall effect in $(B,W)$ plane. Note that for disorder strength
between $W=2$ to $W=20$, the system is the spin-Hall liquid. }
\label{fig3}
\end{figure}

Fig.2a depicts numerical result for the number of edge-states $N$ as
we vary $B$ and $W$. We observe that the edge-states are gradually
destroyed from the subband edge (measured in the lead 1) to the
subband center when $W$ is increased. From Fig.2a we also observe
that $N=2$ edge-states are more stable against disorder than that of
$N=3$. Fig.2c,d show spin-Hall conductance and spin-Hall conductance
fluctuation, respectively, for $W \le 4$\cite{foot1}. They are
perfectly consistent with the general discussion given above, namely
$G_{sH}$ and $rms(G_{sH})$ are finite for $N=$odd edge-states and in
regions when at least one edge-state is destroyed by disorder.

Fig.3 plots the main result of this work, the low field ``phase
diagram" of MSHE. In the numerical calculations of this ``phase
diagram", we have computed 61 values of $B$, 40 values of $W$ from
$W=0$ to $W=4$, and for each pair of $(B,W)$ we averaged over 1000
impurity configurations. The integers in the ``phase diagram"
indicate the number of edge-states $N$. At weak disorder, there are
three possible states: the $N=$ even edge-state induced spin-Hall
insulator, the SOI dominant $N=$ odd edge-state induced spin-Hall
liquid state, and the Zeeman dominant $N=$ odd edge-state induced
fluctuationless spin-Hall liquid. Since large magnetic field favors
Zeeman term, so in $N=$ odd plateau the SOI dominant spin-Hall
liquid appears first for low magnetic field and crosses over to
Zeeman dominant fluctuationless spin-Hall liquid at higher field. As
$W$ increases, the edge-states become destroyed and the system
enters spin-Hall liquid where $G_{sH} \neq 0$ and $rms(G_{sH}) \neq
0$. A chaotic state of MSHE with $G_{sH}=0$ and $rms(G_{sH}) \neq 0$
is reached when $W$ is increased further. Finally, the system enters
a spin-Hall insulator state where $G_{sH}=0=rms(G_{sH})=0$ at large
enough disorder. While this ``phase diagram" is obtained for a
particular value of Rashba SO interaction $t_{so}$, we have checked
that the general topology is the same for other values. In addition,
the MSHE ``phase diagram" in the $(t_{so},W)$ plane for a fixed
magnetic field has similar features. We have also determined the
phase boundary between the chaotic state of MSHE and spin-Hall
insulator that are shown in Fig.3 with the same
resolution\cite{foot1}.

We have so far focused on linear Rashba SOI. A similar analysis can
be carried out for Dresselhaus SOI by adding a term $\beta_{so}
(\sigma_x {\bar k}_x - \sigma_y {\bar k}_y)$ in Eq.(\ref{eq1}).
It is well known that in the absence of Zeeman energy
one has $I_{sH}^z(\alpha_{so}=0,\beta_{so}) = I^z_{sH}
(\alpha_{so},\beta_{so}=0)$ and $I^z_{sH}(\alpha_{so}=\beta_{so}) = 0$.
Therefore, in the absence of Zeeman energy, the MSHE ``phase diagram"
for Dresselhaus SOI is the same as that of the Rashba SOI. In the presence
of Zeeman energy, our numerical results for Dresselhaus SOI give a
similar ``phase diagram". When both Rashba and Dresselhaus terms are
present, a similar ``phase diagram" is also obtained numerically for
$t_{so}=0.2$ and $t_{so2}=0.4$ ($t_{so2}=\beta_{so}/2a$).

In summary, we have determined the low field ``phase diagram" of
mesoscopic spin-Hall effect. The ``phase diagram" is characterized
by values of $G_{sH}$ and $rms(G_{sH})$ in the $(B,W)$ plane and the
main features include a spin-Hall liquid behavior where both
$G_{sH}$ and $rms(G_{sH})$ are nonzero, and by spin-Hall insulator
behavior where both quantities vanish. Furthermore, the spin-Hall
liquid can be induced by $N=$odd edge-states in weak disorder, and
by destroying edge-states for larger disorder. The spin-Hall
insulator behavior, on the other hand, is induced by $N=$even
edge-states, and by very large disorder. The MSHE ``phase diagram"
is found to be true for both linear Rashba and Dresselhaus SO
interactions.

\acknowledgments
This work was financially supported by RGC grant (HKU 7048/06P) from
the government SAR of Hong Kong. H.G is supported by NSERC of Canada,
FQRNT of Qu\'{e}bec and Canadian Institute of Advanced Research.
Computer Center of The University of Hong Kong is gratefully acknowledged
for the High-Performance Computing assistance.

\noindent{$^{*}$ Present address: Physics Department, Hong Kong
University of Science and Technology, Clear Water Bay, Hong Kong

\noindent{$^\dag$} Electronic address: jianwang@hkusub.hku.hk}



\begin{thebibliography}{99}
\bibitem{foot1}
In our numerics, the spin-Hall conductance is taken as zero if
$G_{sH}<0.002 e/4\pi$. The same criterion is applied to the
spin-Hall conductance fluctuation.
\end{thebibliography}
\end{document}